\magnification=1200
\overfullrule=0pt 
\parindent=.8truecm
\def\square{{\vcenter{\vbox{\hrule height .6pt\hbox{\vrule width .6pt 
height 6pt\kern 6pt\vrule width .6pt}\hrule height .6pt}}}}
\hsize=16truecm
\vsize=24truecm 
\def\nt{\noindent}

\def\ss{\vskip1pt}
\def\ms{\vskip3.3pt}
\def\bs{\vskip5pt}

\def\bul{^{\bullet}_{\bullet}}

\def\R{\hbox{\bb R}}
\baselineskip24truept

 at 7truept

\font\BIG=cmbx10 at 14pt

%\font\small=cmr10 at 10pt

\font\bb=msbm10
\def\Z{\hbox{\bb Z}}

\def\C{\hbox{\bb C}}

\def\i#1/#2\par{\item{\hbox to .8truecm{#1\hss}}#2}
\def\ii#1/#2/#3\par{\advance\parindent by .8truecm
       \item{\hbox to .8truecm{#1\hss}
         \hbox to 0.6truecm{#2\hss}}#3
          \advance\parindent by -.8truecm}
\def\iii#1/#2/#3/#4\par{\advance\parindent by 1.6truecm
        \item{\hbox to 0.6truecm{#1\hss}
              \hbox to 0.6truecm{#2\hss}
              \hbox to 0.6truecm{#3\hss}}#4
        \advance\parindent by -1.6truecm}

\bs
\bs
\bs
\ss
\baselineskip24truept
\bs

\bs
\bs
\bs
\ss
\baselineskip24truept
\centerline {\BIG GAUGE SPINORS AND STRING DUALITY}
\bs
\bs
\bs
\bs
\centerline{Brett McInnes*}
\centerline{Department of Mathematics}
\centerline{National University of Singapore}
\centerline{10 Kent Ridge Crescent,} 
\centerline{Singapore 119260}
\centerline{Republic of Singapore}
\bs
\vskip2cm
\nt {\bf ABSTRACT}:  When the gauge groups of the two heterotic string theories are broken, over tori, to their ``$SO(16) \times SO(16)$" subgroups, the winding modes correspond to representations which are spinorial with respect to those subgroups.  Globally, the two subgroups are isomorphic neither to $SO(16) \times SO(16)$ {\it nor to each other}.  Any attempt to formulate the T-duality of the two theories on topologically non-trivial compactification manifolds must therefore take into account various generalizations of the ``spin structure" concept.  We give here a global formulation of T-duality, and show, with the aid of simple examples, that two configurations which {\it appear} to be T-dual at the local level can fail to be globally dual.
\bs
\vskip1cm
\nt PACS : 11.15.Kc, 11.25.-w
\ms
\nt Keywords : Duality, Spin Structures
\vskip2.8cm
\nt * matmcinn@nus.edu.sg

\bs
\bs
\bs
\ss
\pageno=1
\nt {\bf 1.  INTRODUCTION}
\ms
A Riemannian structure on a smooth manifold $M$ is a ${\it reduction}$ [1] of its bundle of linear frames to an $O(n)$ bundle $O(M)$, where $O(n)$ is the group of $n \times n$ orthogonal matrices.  This group is neither connected nor simply connected; however, it can in effect be replaced by simpler groups if $M$ satisfies certain topological conditions.  If $O(M)$ reduces to a bundle with a connected structural group, then $M$ acquires an {\it orientation}, and if this connected group lifts to its universal cover, then $M$ is said to have a {\it spin structure}.  These simplifications are of course not always possible : the $O(4)$ associated with the real projective space $\R P^4$ cannot be reduced to $SO(4)$, and the $SO(4)$ associated with the complex projective space $\C P^2$ cannot be lifted to its non-trivial double cover $Spin(4)$.
\bs
Gauge theory can be regarded as a generalisation of Riemannian geometry, in which one considers arbitrary principal $G$-bundles [1] on $M$, instead of $O(M)$.  As in the Riemannian case, $G$ need be neither connected nor simply connected, and so one might expect to confront problems analogous to the existence of orientations and spin structures.  In fact, this analogy has not attracted a great deal of attention, partly because gauge theory itself does not accord preference to any one of the Lie groups with a given Lie algebra.  If we had reason to believe, for example, that the gauge group of $``SO(10)$" grand unification is ``really" $O(10)$, then indeed it would be of interest to investigate $O(10)$ gauge configurations (over necessarily topologically non-trivial base manifolds) which cannot break to $SO(10)$, and $SO(10)$ configurations which cannot lift to $Spin(10)$.  (Such configurations do exist.)  The point, however, is that we have, a priori, no reason to prefer $O(10)$ to $SO(10)$ or $Spin(10)$ or $Pin(10)$ or any of the other [2] Lie groups with this Lie algebra.  This is where the analogy with Riemannian geometry appears to break down.
\bs
The advent of string theories [3], however, has changed this situation.  These theories are so severely constrained that they do in fact dictate the precise global structures of their gauge groups.  In the case of the ``$E_8 \times E_8$" heterotic theory, the gauge group is $(E_8 \times E_8) \triangleleft \Z_2$, where $\triangleleft$ denotes a semi-direct product, and $\Z_2$ acts by exchanging the two $E_8$ factors.  (See, for example, Ref. $[4]$ for a discussion of the significance of this $\Z_2$.)  Thus, the gauge group is disconnected, and so a property analogous to {\it orientability} must be considered.  However, $E_8 \times E_8$ is simply connected, so there is no analogue of spin structures here.  By contrast, the full gauge group of the ``$SO(32)$" heterotic theory is the group usually denoted by $Spin(32)/\Z_2$.  Unlike $E_8 \times E_8$ (and $SO(32))$ this group has no outer automorphism, and so [5] it has no non-trivial disconnected version; thus, questions of ``orientability" do not arise here.  However, $Spin(32)/\Z_2$ is obviously not simply connected, so the possible non-existence of ``spin structures" is an issue [6].
\bs
We see, then, that the two heterotic string theories do require us to consider the consequences of having topologically non-trivial gauge groups.  The two theories appear, however, to behave in opposite ways, with one gauge group being disconnected but with a simply connected identity component, while the other is connected but not simply connected.  That such appearances are deceptive is, of course, the lesson of the ``duality revolution" [7].  The objective of this work is to understand  the role of these two specific kinds of topological non-triviality in maintaining (or obstructing) the T-duality between the two heterotic string theories.  (Note that, throughout this work, we interpret ``T-duality" in a very broad sense.  {\it Any mapping of the gauge and matter fields of one heterotic theory to those of the other will be called T-duality here}.)
\bs
The two heterotic theories have, of course , different gauge groups, and these groups lead, as above, to different topological complications.  The two Lie {\it algebras}, however, have much in common : in particular, they have a common maximal, maximal-rank subalgebra isomorphic to the algebra of $SO(16) \times SO(16)$.  The $T$-duality between the two heterotic theories has to be established through $(E_8 \times E_8) \triangleleft \Z_2$ and $Spin(32)/\Z_2$ configurations such that each group is broken to the ``common" $SO(16) \times SO(16)$ subgroup.  The problem here, as was pointed out in Ref.[5], is that this ``common" subgroup {\it does not exist} $-$ the two $SO(16) \times SO(16)$ algebras exponentiate to completely different subgroups.  These are
$$[(Spin(16)/\Z_2) \times (Spin(16)/\Z_2)] \triangleleft \Z_2 \to (E_8 \times E_8) \triangleleft \Z_2$$
and
$$[(Spin(16) \times (Spin(16))/(\Z_2 \times \Z_2)] \triangleleft \Z_2 \to Spin(32)/\Z_2,$$
where, in both cases, the semidirect product with $\Z_2$ is defined through the exchange of the two local factors.  Notice that the subgroups are certainly more closely related than the original pair : they have the same Lie algebra, they both have two connected components, and they both have $\Z_2 \times \Z_2$ as the fundamental group of the identity component; but they are not isomorphic.  However, it was also shown in Ref. [5] that they have a common double cover, which $-$ crucially $-$ is {\it not} the universal cover.  $T$-duality, therefore, can only be established at that level.  {\it This assumes, however, that the relevant gauge configurations obtained from breaking} $(E_8 \times E_8) \triangleleft \Z_2$ {\it and} $Spin(32)/\Z_2$ {\it can indeed be lifted to this common double cover}.  This, again, involves a gauge-theoretic analogue of the question of the existence of spin structures.  Clearly, a full understanding of $T$-duality will require an analysis of the theory of {\it gauge spinors} for these groups.  It will also be useful to have concrete examples of $(E_8 \times E_8) \triangleleft \Z_2$ and $Spin(32)/\Z_2$ configurations such that the above subgroups do {\it not} lift to the double cover : cases in which $T$-duality is topologically obstructed.
\bs
We begin with a general analysis of the problem of comparing two distinct gauge theories over a given manifold.  Then we turn to the details of the specific groups involved in string theory, before giving a general formulation of the topological aspects of $T$-duality.  We conclude with several very simple concrete examples which show that it is necessary to take certain obstructions into account when discussing  $T$-duality over topologically non-trivial space-times.
\bs
\ms
\nt {\bf 2.  GAUGE GROUPS WITH ``COMMON" SUBGROUPS}.

\ms
Let $G_1$ and $G_2$ be compact Lie groups such that both have a proper compact subgroup isomorphic to $H$.  (For the present, we assume that the isomorphisms are global; we shall consider the local case below.)  $T$-duality involves, among other things, a process of comparing or exchanging gauge and matter fields associated with two different gauge groups.  It is not entirely clear what this notion means from a global point of view, however.  Gauge and matter fields associated with $G_1$ are essentially defined not on the space-time $M$, but rather on some principal $G_1$ fibre  bundle $P_1$ over $M$, and similarly for $G_2$.  Evidently it does not make sense in general to speak of ``comparing" fields on $P_1$ with fields on $P_2$; to make sense of it, we must exploit the common subgroup $H$.
\bs
Let us begin with a principal $G_1$ bundle $P_1$ which reduces to an $H$-subbundle $Q$.  (Notice that this is a condition on $P_1$, and that the condition is more restrictive for ``smaller" 
$H : $ in the extreme case where $H = \{identity \}$, $P_1$ must be trivial.  Thus the following construction is most satisfactory if $H$ is maximal in $G_1$ and $G_2$.)  Now, because $H$ is also a subgroup of $G_2$, it acts on $Q \times G_2$ by
$$h \in H : (q,g_2) \to (qh, h^{-1}g_2),$$
and so we can define
$$P_2 = (Q \times G_2)/ H.$$
Then $P_2$ admits a natural action to the right by $G_2$, with $P_2/G_2 = M$, and one can show [8] that $P_2$ is a principal $G_2$-bundle over $M$.  Thus any $G_1$-bundle which reduces to $H$ will define a $G_2$ bundle in a natural way.

Now let ${\cal G}_i$ be the Lie algebra of $G_i$ and let $\cal H$ be the algebra of $H$.  Since $H$ is assumed to be compact, a standard procedure allows us to construct a direct sum,
$${\cal G}_i = {\cal H} \oplus {\cal B}_i,$$
such that $Ad_{G_{i}}(H)$, the restriction to $H$ of the adjoint representation of $G_i$, satisfies $Ad_{G_{i}}(H){\cal B}_i = {\cal B}_i$.  Therefore if $\omega_1$ is any connection form on $P_1$ and $\omega_1 |_{_Q}$ is its restriction to $Q$, then setting
$$\omega_1|_{_Q} = {\cal H} \omega_1|_{_Q} \oplus {\cal B}_1\omega_1|_{_Q},  $$
we find that each component transforms separately under the action of $Ad_{G_{1}}(H)$.  It follows [1] that ${\cal H} \omega_1 |_{_Q}$ defines a connection on $Q$.  Now let $P_2$ be defined as above, and let $[q,g_2]_H \in P_2$, where $q \in Q, g_2 \in G_2$.  Then we can define a connection on $P_2$ by
$$\omega_2([q,g_2]_H) = Ad_{G_{2}}(g^{-1}_2) {\cal H} \omega_1(q);$$
it is easy to verify that $\omega_2$ is well-defined on $P_2$ and that it is a connection form.  Thus any connection on $P_1$ defines a connection on $P_2$.  (Adding a tensorial ${\cal B}_2$-valued form [1] to $\omega_2$ will give us other connections on $P_2$ with $\cal H$-components coinciding with the $\cal H$-component of $\omega_1$)
\bs
Next, consider matter fields.  These are conventionally thought of as cross-sections of associated bundles, but for purposes of global geometry it is preferable to regard them as equivariant functions on principal bundles.  Let $f$ be a function on a principal $G$-bundle $P$, taking its values in a finite-dimensional vector space $V$ which affords a representation $\rho$ of $G$.  If $f$ satisfies $R^*_gf = \rho(g^{-1})f$ for every $g \in G$, then [1] $f$ defines a cross-section of the associated bundle defined by $V$ and $\rho$; conversely, every such cross-section defines a function $f$.  We shall therefore refer to such functions on $P$ as matter fields.  (The reader who finds this unnatural may wish to think of the ``traditional" definitions of vector fields and matter multiplets in terms of components.  The definition of matter fields used here is simply a global version of that tradition.)  Let $f_1$ be a matter field on the bundle $P_1$ discussed earlier, taking its values in $V_1$ and equivariant with respect to $\rho_1$.  Restricting $f_1$ to $Q$ and $\rho_1$ to $H$, we obtain a matter field on $Q$ taking its values in $V_1$.  The Peter-Weyl theorem [9] implies, since all of our groups are compact, that there exists a representation of $G_2$, $\rho_2$, which restricts to this representation of $H$.  We can now define a matter field $f_2$ on $P_2$ by
$$f_2([q,g_2]_H) = \rho_2(g^{-1}_2)f_1(q);$$
this is well-defined, since $\rho_2((h^{-1}g_2)^{-1}) = \rho_2(g^{-1}_2)\rho_1(h)$ when acting on $V_1$.  (See also Ref. [9] on Frobenius reciprocity.)  Thus we see that $G_1$ matter fields can be re-interpreted as $G_2$ matter fields.
\bs
To conclude, then, we have made the following two observations.  First, if $G_1$ and $G_2$ are gauge groups, then in general it is difficult to compare $G_2$ gauge and matter fields with $G_1$ gauge and matter fields, because these fields ``live on" different spaces.  Second, however, such comparisons {\it are} possible if $G_1$ and $G_2$ have a large common subgroup $H$, provided that the $G_1$ and $G_2$ principal bundles reduce to a common $H$-bundle.  One can only expect any kind of ``duality" to work in the latter case.
\bs
\ms
\nt {\bf 3.  GAUGE SPINORS}
\bs
Let $H$ be a connected, compact, non-simply-connected Lie group, and let $\widetilde H$ be some non-trivial finite cover (not necessarily the universal cover) of $H$, so that $\widetilde H$ has a finite central subgroup $N$ with ${\widetilde H}/N = H$.  Let $\widetilde Q$ be a principal $\widetilde H$-bundle over a manifold $M$; then $Q = {\widetilde Q}/N$ is a principal $H$-bundle over  $M$.  If $f$ is a matter field on $\widetilde Q$, then it may or may not be invariant with respect to $N$; this will be determined by the representation in which it takes its values.  We shall say that $f$ is {\it vectorial with respect to} $H$ if it is invariant with respect  to $N$, and {\it spinorial with respect to} $H$ otherwise.  Clearly $f$ descends to a matter field on $Q$ if and only if it is vectorial.
\bs
In Grand Unification [10], a given gauge group (usually the ``standard" group) $H$ is regarded as a subgroup of some larger group $G$.  The objective is to gain some control over the kinds of matter fields in the theory : only those representations of $H$ which can be regarded as restrictions of $G$-representations are accepted.  Of course, by the Peter-Weyl theorem, all $H$-representations are restrictions of {\it some} $G$-representation, but in practice one requires the $G$-representation to be ``small".  {\it In addition}, however, grand unified theories have the effect of {\it ruling out} $H$ {\it gauge spinors}.  For if $\rho$ is a representation of $\widetilde H$ which does not  descend to a representation of $H$, then there is no sense in which $\rho$ can be said to arise from any representation of $G$.  For example, the unbroken gauge group of the standard model [11] is the unitary group $U(3)$, which is triply covered by $U(1) \times SU(3)$.  The usual representations [11] of $U(3)$ appearing in the standard model can all be obtained from (``small") $SU(5)$ representations, but the opposite is true of those (``spinorial") representations of $U(1) \times SU(3)$ which do not descend to representations of $U(3)$.  The inclusion of matter fields defined by representations of that kind {\it would defeat the whole point of grand unification.}
\bs
Now it is essential to understand that the gauge groups of (heterotic) string theories {\it are not grand unification groups in the usual sense}; ``unification" in string theory is much more subtle.  Consider, for  example, either $E_8$ in $(E_8 \times E_8) \triangleleft \Z_2$.  This group has a maximal, maximal-rank subgroup globally isomorphic to $Spin(16)/\Z_2$.  In the fermionic formulation of the $E_8$ theory, one has sixteen left-handed fermions transforming as the ``{\bf 16} of $Spin(16)/\Z_2$".  However, no such representation exists $-$ the {\bf 16} is a representation of $Spin(16)$, but {\it not} of $Spin(16)/\Z_2$.  In other words, these fields are spinorial with respect to $Spin(16)/\Z_2$.  This kind of behaviour, which cannot occur in a true $E_8$ grand unified theory, is of vital importance in string theory.  Most importantly for our purposes, when the heterotic theories are compactified on tori or toral orbifolds, the {\it winding modes} [3] which play a crucial role in T-duality are in fact gauge spinors with respect to certain groups.  We shall return to this point below.
\bs
In view of these remarks, we need to modify the formalism of the preceding section to allow for the possibility that $G_1$ and $G_2$ have subgroups which are locally but {\it not} globally isomorphic.  That is, the respective subgroups, $H_1$ and $H_2$, are {\it not} isomorphic, but they have the same Lie algebra.  Then there exists a group $H^\ast$ with finite central subgroups $N_1$ and $N_2$, such that $H_i = H^\ast/N_i$.  (Note that $H^\ast$ is {\it not} a subgroup of either $G_1$ or $G_2$, and that it need not be simply connected.)  Suppose as before that $P_1$ is a principal $G_1$ bundle admitting an $H_1$ subbundle $Q_1$.  {\it Assume} now that $Q_1$ has a non-trivial $N_1$ cover $Q^\ast$ which  is an $H^\ast$ bundle over the same base.  Then $Q_2 = Q^\ast/N_2$ is a well-defined $H_2$ bundle, and $Q_2$ can be extended as usual to a $G_2$ bundle $P_2$.  We have a pair of bundles reducing to subbundles with a {\it common} gauge spinor bundle.
\bs
Next, let $\omega_1$ be a connection on $P_1$, and, as in the preceding section, let $\phi _1 = {\cal H}\omega_1|_{Q_{1}}$ be the corresponding connection on $Q_1$.  Pull-backs of connections do {\it not} always define connections, but, if $\pi$ denotes the projections $\pi : Q^\ast \to Q_1$, $\pi : H^\ast \to H_1$, we have, for any $h \in H^\ast$,
$$\eqalign{
 R^\ast_h\pi ^\ast\phi_1 &= (\pi \circ R_h)^\ast \phi_1 = \pi^\ast R^{\ast}_{\pi (h)}\phi_1 \cr
                         &= Ad(\pi (h))\pi^\ast \phi_1 = Ad(h) \pi^\ast \phi_1,\cr}$$
\ms
since $N_1$ is central.  Thus $\pi^\ast \phi_1$ is indeed a connection on $Q^\ast$; it projects naturally to a connection on $Q_2$, and from there it can be extended to $P_2$ as usual.  In short, gauge fields can be freely moved, via $Q^\ast$, from one side to the other.  Matter fields, however, are different.
\bs
Notice first that {\it all} matter fields on both $Q_1$ and $Q_2$ can be pulled back to $Q^\ast$, which provides an arena for comparing or exchanging such fields.  (This is why it is more convenient to regard matter fields as equivariant functions on principal bundles rather than as cross-sections.)  However, we may wish to consider matter fields on $Q^\ast$ which are not pulled back from either $Q_1$ or $Q_2$; these will be spinorial with respect both to $H_1$ and to $H_2$.  Notice too that fields on $Q^\ast$ which are vectorial with respect to $H_1$  can be spinorial with respect to $H_2$, and vice versa; {\it the status of a given field can depend on one's point of view}.  This ``relativity" is an important ingredient of $T$-duality.  In general, fields on $Q^\ast$ fall into four sectors : those which are invariant with respect to both $N_1$ and $N_2$, those which are invariant with respect to one but not the other, and those which are invariant with respect to neither.
\bs
Let us now turn to the specific groups involved in heterotic string theory.
\bs
\ms
\nt {\bf 4.  A GLOBAL FORMULATION OF T-DUALITY}
\bs
The gauge groups of the heterotic theories are $(E_8 \times E_8) \triangleleft \Z_2$ and $Spin(32)/\Z_2$.  Both of these have subgroups locally isomorphic to $SO(16) \times SO(16)$, but the global picture is not so simple.
\bs
The spin group $Spin(4m)$ has a centre of the form [12]
$$\Z_2 \times \Z_2 = \{\pm 1, \pm  {\hat K}_{4m}\},$$
where ${\hat K}_{4m}$ covers the $SO(4m)$ matrix  $-I_{4m}$.  Of course, we have $SO(4m) = Spin(4m)/\{\pm 1\}$, but since $({\hat K}_{4m})^2 = 1$, we can  also define $Spin(4m)/\{1, {\hat K}_{4m}\}$, which differs from $SO(4m)$ for all $m \not= 2$.  This is the group usually called $Spin(4m)/\Z_2$, but this notation would be too confusing for our purposes : we shall call it $Semispin (4m)$ instead.  The name is motivated by the fact that these are the groups for which the {\it half-spin} representations [12] are faithful; the fundamental half-spin representation of $Semispin(4m)$ is $(2^{2m-1})$-dimensional.
\bs
As was mentioned earlier, $E_8$ contains $Semispin(16)$, not $SO(16)$ or $Spin(16)$. (The adjoint of $E_8$ decomposes as  ${\bf 248} = {\bf 128} \oplus {\bf 120}$, where the $\bf 128$ is the half-spinor of $Semispin(16)$, and $\bf 120$ is its adjoint.)  Thus $(E_8 \times E_8) \triangleleft \Z_2$ contains $(Semispin(16) \times Semispin(16)) \triangleleft \Z_2$.  To see that the $\Z_2$ factor must be included, note that, for a discussion of T-duality on a torus, we must break $(E_8 \times E_8) \triangleleft \Z_2$ by a Wilson loop; the latter breaks a gauge group to the centraliser of some element of the group.  Now $Semispin(16)$ is the centraliser, in $E_8$, of $(-1)^\ast$, the projection of $-1 \in Spin(16)$.  Since $((-1)^\ast, (-1)^\ast)$ is clearly invariant under the exchange automorphism, we see that the centraliser of this element of $(E_8 \times E_8) \triangleleft \Z_2$ is indeed $(Semispin(16) \times Semispin(16)) \triangleleft \Z_2$.
\bs
One might reasonably expect $Semispin(32)$, the other heterotic gauge group, to contain this same subgroup, just as $SO(32)$ contains $SO(16) \times SO(16)$.  {\it This is not the case, however}.  Note first that $Spin(32)$ does behave in the expected way : it contains $Spin(16) \bullet Spin(16)$, where the dot means that $-1$ in one $Spin(16)$ is identified with $-1$ in the other.  However, upon projecting $Spin(32)$ to $Semispin(32)$, we obtain [5] the subgroup
$$Spin(16) \bul Spin(16) = [Spin(16) \times Spin(16)]/(\Z_2 \times \Z_2),$$ in which the entire centre of one $Spin(16)$ is identified with that of the other.  (We shall denote the elements of this group by $A \bul B$, for $A, B \in Spin(16)$.)  Again, for Wilson-loop breaking on a torus we wish to express $Spin(16) \bul Spin(16)$ as a centraliser, but this cannot be done; instead, the $Semispin(32)$ centraliser of ${{\hat K}}_{{16}} {\bul } 1$ is isomorphic to 
$$(Spin(16) \bul Spin(16)) \triangleleft \Z_2,$$
where $\Z_2$ acts by exchanging the two local $Spin(16)$ factors [5].  Thus we see that the relevant subgroups of $(E_8 \times E_8) \triangleleft \Z_2$ and $Semispin(32)$ are {\it locally but not globally isomorphic}.
\bs
Before moving on, let us note the following.  The groups $(E_8 \times E_8) \triangleleft \Z_2$ and $Semispin(32)$ exhibit topological complications of two very different kinds : one is disconnected, the other non-simply-connected.  Yet their ``$SO(16) \times SO(16)$" subgroups are much more similar.  This similarity, which is so crucial to T-duality, is however very delicately dependent on the differences between $Spin(4m)$ and $Semispin(4m)$.  On the one hand, $E_8 \times E_8$ contains a non-simply-connected subgroup {\it because} $E_8$ contains $Semispin(16)$ and {\it not} $Spin(16)$.  On the other hand, the other subgroup is disconnected {\it because} the gauge group is $Semispin(32)$ and not $Spin(32)$.  By this we mean the following : a theorem of Bott $[13]$ states that the centraliser of any element of a connected, compact, {\it simply connected} Lie group must be {\it connected}.  Indeed, the centraliser of ${\hat K}_{16} \bullet 1$ in $Spin(32)$ is the {\it connected} group $Spin(16) \bullet Spin(16)$.  In this sense, $(Spin(16) \bul Spin(16)) \triangleleft \Z_2$ is disconnected (like $(Semispin(16) \times Semispin(16)) \triangleleft \Z_2)$ {\it because} $Semispin(32)$ is {\it not} simply connected.  Thus the apparently opposite topological complications of the two heterotic gauge groups are unified by T-duality.  This general fact was made explicit in the case of the CHL compactifications [14] by the work of Lerche et al [4] who show that the moduli space of flat connections in the $T^2$ compactification of the $Semispin(32)$ heterotic string has two connected components ({\it because} $Semispin(32)$ is not simply connected), and so does the moduli space of flat connections in the $T^2$ compactification of the $(E_8 \times E_8) \triangleleft \Z_2$ string ( {\it because} this group is not connected).  The CHL compactifications on each side then correspond to the extra components of the moduli spaces.  In general, we should always expect phenomena associated with the fact that $Semispin(32)$ is not simply connected to be related in some way to phenomena associated with the fact that $(E_8 \times E_8) \triangleleft \Z_2$ is not connected.  We shall return to this later.
\bs
Let us now apply the formalism of the preceding section to  the case at hand.  Our objective is to construct a $Semispin(32)$ bundle $P_1$ admitting a $(Spin(16) \bul Spin(16)) \triangleleft \Z_2$ subbundle $Q_1$, such that $Q_1$ lifts to an $H^\ast$ bundle $Q^\ast$, which in turn projects to  a $(Semispin (16) \times Semispin(16)) \triangleleft \Z_2$ bundle $Q_2$.  Our first objective is of course to identify the group $H^\ast$.  For simplicity, let us temporarily dispense with the $\Z_2$ factors.
\bs
Since $Spin(16) \bul Spin(16)$ and $Semispin(16) \times Semispin(16)$ are locally isomorphic, they have the same universal covering group, namely $Spin(16) \times Spin(16)$.  This is definitely {\it not} a good choice for $H^\ast$, however, both for mathematical reasons and, more importantly, physical ones.  Let $Q_1$ be a principal $Spin(16) \bul Spin(16)$ bundle admitting a non-trivial $\Z_2 \times \Z_2$ cover $Q^\ast$ which is a $Spin(16) \times Spin(16)$ bundle over the same base.  Then $Q^\ast/\{(1,1), (-1,-1) \}$ is a  $Spin(16) \bullet Spin(16)$ bundle, and so, $Spin(16) \bullet Spin(16)$ being a subgroup of $Spin(32)$, this bundle can be extended to a $Spin(32)$ bundle.  Thus, if we take $H^\ast = Spin(16) \times Spin(16)$, we are ruling out any $Semispin(32)$ configuration which fails to lift to a $Spin(32)$ configuration.  However, it is known that T-duality cannot work without $Semispin(32)$ gauge fields of this kind : Berkooz et al [6] show that the DMW models [15] for the $(E_8 \times E_8) \triangleleft \Z_2$ theory are T-dual to such $Semispin(32)$ configurations.  Clearly it is necessary to find some $H^\ast$ which is a {\it double} (rather than quadruple) cover of both $Semispin(16) \times Semispin(16)$ and $Spin(16) \bul Spin(16)$.  Notice that it is not obvious that such a group exists $-$ for example, $Semispin(16) \times Semispin(16)$ and $SO(16) \times SO(16)$ have no common cover other than $Spin(16) \times Spin(16)$.
\bs
There are in fact no fewer than 22 distinct connected groups locally isomorphic to $Spin(16) \times Spin(16)$.  Of these, however, only five have fundamental groups isomorphic to $\Z_2$, namely $SO(16) \times Spin(16)$,
$Semispin(16) \times Spin(16)$, $Spin(16) \bullet Spin(16)$, a certain  group defined by identifying $-1$ in one $Spin(16)$ with ${\hat K}_{16}$ in another, and $Spin(16) \ast Spin(16)$, defined by identifying ${\hat K}_{16}$ in one $Spin(16)$ with ${\hat K}_{16}$ in  the \hfil\break other : that is,
$$Spin(16) \ast Spin(16) = (Spin(16) \times Spin(16))/\{(1,1), ({\hat K}_{16}, {\hat K}_{16})\}.$$  It is  not difficult to show that none of the first four candidates covers both $Semispin(16) \times Semispin(16)$ and $Spin(16) \bul Spin(16)$.  But we have
$$Semispin(16) \times Semispin(16) = (Spin(16) \ast Spin(16))/\{(1 \ast1), ({\hat K}_{16} \ast 1\},$$
$$Spin(16) \bul Spin(16) = (Spin(16) \ast Spin(16))/\{(1\ast 1), (-1) \ast (-1)\}.$$
Thus, $Spin(16) \ast Spin(16)$ double-covers both groups, and it is the unique (connected) physically acceptable candidate for $H^\ast$.  (As $Spin(16) \ast Spin(16)$ is symmetrical between the two factors, we can define $(Spin(16) \ast Spin(16)) \triangleleft \Z_2$ in the obvious way; and since ${\hat K}_{16} \ast 1$ and $(-1) \ast (-1)$ are invariant under the action of $\Z_2$, it is clear that  $(Spin(16) \ast Spin(16)) \triangleleft \Z_2$ double-covers $(Semispin(16) \times Semispin(16)) \triangleleft \Z_2$ and $(Spin(16) \bul Spin(16)) \triangleleft \Z_2.)$
\bs
The framework for any global formulation  of the $T$-duality between the heterotic string theories is therefore as follows.  We can begin with a $Semispin(32)$ bundle $P_1$ admitting a $(Spin(16) \bul Spin(16)) \triangleleft \Z_2$ subbundle $Q_1$.  We {\it assume} that $Q_1$ has a non-trivial double cover  $Q^\ast$ which is a $(Spin(16) \ast Spin(16)) \triangleleft \Z_2$ bundle over the same base.  On $Q^\ast$ we will have certain matter fields, including winding modes which may be gauge spinors from the $Semispin(32)$ point of view.  But let $Q_2 = Q^\ast/\{1 \ast 1, {\hat K}_{16} \ast 1 \}$.  Then $Q_2$ is a $(Semispin(16)) \times (Semispin(16)) \triangleleft \Z_2$ bundle over the same base as $Q_1$, and $Q_2$ may be regarded as a subbundle of an $(E_8 \times E_8) \triangleleft \Z_2$ bundle $P_2$.  The matter fields on $Q^\ast$ can again be interpreted as vectorial or spinorial $(E_8 \times E_8) \triangleleft \Z_2$ matter fields.  The crucial point is that all of the matter (and gauge) fields ``live" on the same manifold, $Q^\ast$, and it therefore makes sense to speak of mapping  one to another under $T$-duality.  Furthermore, since the status of a given matter field on $Q^\ast$ (as a gauge vector or a gauge spinor) is dependent on whether one takes the $Semispin(32)$ or the $(E_8 \times E_8) \triangleleft \Z_2$ point of view, the characteristic exchange of winding with non-winding modes can be easily accommodated.  Finally, the assumption that $Q^\ast$ exists, to which we must of course return, in no way requires $P_1$ to lift to a $Spin(32)$ bundle, and so we can handle $DMW$-like configurations.
\bs
Some concrete examples may be helpful.  The adjoint of $Semispin(32)$ coincides with that of $SO(32)$, which decomposes under $SO(16) \times SO(16)$ as ${\bf 496} = ({\bf 16}$, ${\bf 16}) \oplus ({\bf 120}, {\bf 1}) \oplus ({\bf 1}$, ${\bf 120})$, where ${\bf 120}$ is the adjoint of $SO(16)$.  The tensor product $({\bf 16}, {\bf 16})$ is a legitimate representation of $Spin(16) \bul Spin(16)$, though $({\bf 16}$, ${\bf 1}) \oplus ({\bf 1}$, ${\bf 16})$ would not be; for $(-I_{16}, -I_{16}) \in SO(16) \times SO(16)$ is in the kernel of $({\bf 16}, {\bf 16})$ but not that of $({\bf 16}, {\bf 1}) \oplus ({\bf 1}, {\bf 16})$, while ${{\hat K}_{16}} {\bul} {{\hat K}_{16}}$, the corresponding element of $Spin(16) \bul Spin(16)$, equals the identity.  A matter field $f_1$ on $Q^{\ast}$ taking its values in  $({\bf 16}, {\bf 16})$ is therefore a gauge vector from the $Semispin(32)$ point of view.  On the other hand, a matter field $h_1$ on $Q^\ast$ taking its values in the half-spin representation $({\bf 128}, {\bf 1}) \oplus ({\bf 1}, {\bf 128})$ (which is a legitimate representation of $Spin(16) \ast Spin(16))$ will satisfy
$$h_1 \circ R_{(-1)\ast(-1)} = - h_1,$$
where $R_g$ denotes the action of $g \in Spin(16) \ast Spin(16)$ on $Q^\ast$, so $h_1$ is a gauge spinor for $Semispin(32)$; such fields will arise as winding modes when the corresponding string theory is compactified on a torus.  From the $E_8 \times E_8$ side, we can also have a gauge {\it vector} $h_2$ taking its values in $({\bf 128}, {\bf 1}) \oplus ({\bf 1}, {\bf 128})$ (this being a legitimate representation of $Semispin(16) \times Semispin(16))$ and a winding mode $-$ a gauge {\it spinor} $-$ $f_2$, taking its values in $({\bf 16}, {\bf 16})$ (which is a representation of $Spin(16) \ast Spin(16)$ but {\it not} of $Semispin(16) \times Semispin(16))$.  In addition, we can have matter fields $g_1, g_2$ from both sides, taking values in $({\bf 120}, {\bf 1}) \oplus ({\bf 1}, {\bf 120})$; as this is a representation  of {\it both} $Spin(16) \bul Spin(16)$ {\it and} $Semispin(16) \times Semispin(16)$, $g_1$ and $g_2$ are gauge vectors from both points of view.  (The same would be true of matter fields, if any, taking values in $({\bf 128}, {\bf 128})$.)  Then $T$-duality acts by exchanging $f_1 \leftrightarrow f_2$, $g_1 \leftrightarrow g_2$, $h_1 \leftrightarrow h_2$, and so on.  The details need not concern us here : the point is that we now have a meaningful global formulation of $T$-duality.  That is, the fields being exchanged ``live on" the same manifold, $Q^\ast$, and they take values in the same representations $-$ despite the fact that some of these representations do not make sense for the original groups obtained by Wilson-loop breaking of $(E_8 \times E_8) \triangleleft \Z_2$ and $Semispin(32)$.  The fact that the respective ``$SO(16) \times SO(16)$" subgroups are {\it not} isomorphic leads to an ambiguity in the vector/spinor status of matter fields $-$ but this ambiguity proves to be nothing but a facet of $T$-duality.
\bs
One question remains, however : does $Q^\ast$ actually exist?
\bs
\nt {\bf 5.  EXISTENCE OF GAUGE SPIN STRUCTURES.}

\bs
The $T$-duality of the two heterotic theories was originally formulated on $\R^9 \times S^1$, where $S^n$ denotes the $n$-sphere.  On this manifold, every $Spin(16) \bul Spin(16)$ principal bundle lifts to a $Spin(16) \ast Spin(16)$ bundle, and so does every $Semispin(16) \times Semispin(16)$ bundle.  The programme outlined in the preceding section can therefore be completed : every $Spin(16) \bul Spin(16)$ configuration is $T$-dual to some $Semispin(16) \times Semispin(16)$ configuration over $\R^9 \times S^1$.
\bs
Recently, attempts have been made (most notably in Ref.[6]) to formulate $T$-duality on more complicated manifolds, particularly $K3$.  Evidently, the above lifting problems will be more complex in these cases.
\bs
Let $P$ be any $(Spin(16) \bul  Spin(16)) \triangleleft \Z_2$ bundle over a 
manifold $M$.  Then the fibration $(Spin(16) \bul Spin(16)) \triangleleft \Z_2 \longrightarrow P \longrightarrow M$ yields the exact sequence [12]
$$0 \to H^0(M, \Z_2) \to H^0(P,\Z_2) \to H^0((Spin(16) \bul Spin(16)) \triangleleft \Z_2, \Z_2) \to H^1(M,\Z_2),$$
where $H^i(X,\Z_2)$ is the ith cohomology group of $X$ with $\Z_2$ coefficients.  Clearly \hfil\break $H^0((Spin(16) \bul Spin(16)) \triangleleft \Z_2, \Z_2)$ has a non-trivial element; let us call it $J^\ast_{16}$.  If $f_P$ is the last homomorphism in the exact sequence, we define
$$x_1(P) = f_P(J^\ast_{16}) \in H^1(M,\Z_2).$$
If (and only if) $x_1(P) = 0$, then by the exactness of the above sequence, $H^0(P,\Z_2) \not= 0$, that is, $P$ is disconnected.  Then  $P$ can be reduced to a $Spin(16) \bul Spin(16)$ subbundle.  This is  of course the analogue of orientability for this group.
\bs
Suppose that $x_1(P) = 0$, so that, in effect, $P$ is a $Spin(16) \bul Spin(16)$ bundle.  The fibration $Spin(16) \bul Spin(16) \to P \to M$ yields [12] another exact sequence,
$$0 \to H^1(M,\Z_2) \to H^1(P,\Z_2) \to H^1(Spin(16) \bul Spin(16), \Z_2) \to H^2(M,\Z_2).$$
Now $Spin(16) \bul Spin(16)$ is defined by factoring $Spin(16) \times Spin(16)$ by its subgroup $\{(1,1), (-1,-1), ({\hat K}_{16},{\hat K}_{16}), (-{\hat K}_{16}, -{\hat K}_{16})\}$.  Hence we can regard this as the fundamental group of  $Spin(16) \bul Spin(16)$, and, since it is isomorphic to $\Z_2 \times \Z_2$, this is also $H^1(Spin(16) \bul Spin(16), \Z_2)$.  Thus, again denoting the last homomorphism in the exact sequence by $f_P$, we define
$$\eqalign{
{\tilde w}_2(P) &= f_P(({\hat K}_{16}, {\hat K}_{16})) \in H^2(M,\Z_2), \cr
x_2(P) &= f_P((-1,-1)) \in H^2(M,\Z_2).\cr}$$ 
(Since $Spin(16) \times Spin(16)$ has an outer automorphism mapping 
$({\hat K}_{16}, {\hat K}_{16})$ to $(-{\hat K}_{16},-{\hat K}_{16})$,  $f_P((-{\hat K}_{16}, - {\hat K}_{16}))$ gives nothing new.)  Now we have the following isomorphisms, in which the notation is self-explanatory.
$$\eqalign{
Spin(16) \bul Spin(16) &= (Spin(16) \bullet Spin(16))/\{1 \bullet 1, {\hat K}_{16} \bullet {\hat K}_{16} \} \cr
&= (Spin(16) \ast Spin(16))/\{1 \ast 1, (-1) \ast (-1) \}.\cr}$$
Furthermore, $H^1(P,\Z_2)$ describes the possible non-trivial double covers of $P$.  The exactness of the above sequence therefore implies that if ${\tilde w}_2(P) = 0$, then $P$ has a non-trivial double cover ${\tilde P}$ which is a $Spin(16) \bullet Spin(16)$ bundle over $M$.  If $P$ is regarded as a subbundle of a $Semispin(32)$ bundle, then the latter will have a non-trivial double cover which is a $Spin(32)$ bundle over $M$.  In such a case one says [6] that $P$ {\it has a vector structure}.  (Notice that the problem is really to lift $Semispin(32)\;to\; Spin(32)$; the projection from $Spin(32)$ to $SO(32)$, which motivates the terminology, can always be done.)  Thus ${\tilde w}_2(P)$ is just the {\it generalised Stiefel-Whitney class} [6] which obstructs the existence of ``vector structures".
\bs
In the same way, $x_2(P) = 0$ means that $P$ has {\it another} non-trivial double cover \hfil\break $P^\ast$ , which is a $Spin(16) \ast Spin(16)$ bundle over $M$.  We shall say in this case that $P$ admits an {\it exceptional structure}, because $P^\ast/\{1 \ast 1, {\hat K}_{16} \ast 1 \}$ is a $Semispin(16) \times Semispin(16)$ bundle which can be extended to an $E_8 \times E_8$ bundle.
\bs
We can now complete our global formulation of heterotic $T$-duality.  We begin with a $Semispin(32)$ bundle $P_1$ admitting a $(Spin(16) \bul Spin(16)) \triangleleft \Z_2$ subbundle $Q_1$.  {\it We require}
$$x_2(Q_1) = 0.$$
Let $Q^\ast$ be a non-trivial double cover of $Q_1$ with structural group $(Spin(16) \ast Spin(16)) \triangleleft \Z_2$.  Then $Q^\ast$ is automatically a non-trivial double cover of a certain $(Semispin(16) \times Semispin(16)) \triangleleft \Z_2$ bundle $Q_2 = Q^\ast/\{1 \ast 1, {\hat K}_{16} \ast 1\}$, which may be regarded as a subbundle of an  $(E_8 \times E_8) \triangleleft \Z_2$ bundle $P_2$.  As explained in the preceding section, $Q^\ast$ is the arena for the exchanges defining $T$-duality.
\bs
Several remarks should be made at this point.  The first and most important is that the condition $x_2(Q_1) = 0$ will not, of course, always be satisfied for all base manifolds $M$.  It {\it is} satisfied for all bundles over $\R^9 \times S^{1}$, since $H^2(\R^9 \times S^1, \Z_2) = 0$, but on more complex manifolds we can certainly find $Semispin(32)$ configurations such that $Q_1$ does not lift to any $Q^\ast$.  In such a case, the configuration has no globally well-defined $E_8 \times E_8$ partner : we can say that {\it there is a topological obstruction to T-duality}.  This is a failure of $T$-duality (see Ref. [16]) which would pass undetected at the local level.  Notice that $x_2$ and ${\tilde w}_2$ take values in the same cohomology group, so one must beware of this possibility in any situation in which the existence of a vector structure is questionable.  On the other hand, the two obstructions are independent : the existence of a vector structure {\it in no way guarantees the existence of an exceptional structure} or vice versa.  (The existence of two obstruction classes is of course due to the greater topological complexity of $Spin(16) \bul Spin(16)$ compared with  $SO(32)$ or $Semispin(32)$.)
\bs
Secondly, we have throughout assumed that one begins with a $Semispin(32)$ bundle and works towards $(E_8 \times E_8) \triangleleft \Z_2$.  However, one is entitled to begin with an \hfil\break $(E_8 \times E_8) \triangleleft \Z_2$ bundle $P_2$ admitting a $(Semispin(16) \times Semispin(16)) \triangleleft \Z_2$ subbundle $Q_2$.  There is an element $y_1(Q_2) \in H^1(M, \Z_2)$ which obstructs the reduction of $Q_2$ to a $Semispin(16) \times Semispin(16)$ bundle.  Assuming that $y_1(Q_2) = 0$ and again calling the reduced bundle $Q_2$, we can ask  whether $Q_2$ lifts to a $Spin(16) \ast Spin(16)$ bundle $Q^\ast$; the obstruction is a certain element $y_2(Q_2) \in H^2(M,\Z_2)$.  If indeed $y_2(Q_2) = 0$, then  $Q^\ast$ exists, and we can define
$$Q_1 = Q^\ast/\{1 \ast 1, (-1) \ast (-1)\},$$
a $Spin(16) \bul Spin(16)$ bundle, and so T-duality goes through.  We stress that ${\tilde w}_2, x_2,$ and $y_2$ are all distinct from each other and from the standard Stiefel-Whitney class $w_2$.  The latter would be relevant to the study of $SO(16) \times SO(16)$ bundles lifting to \hfil\break $Spin(16) \bullet Spin(16)$ bundles; this may not seem to be directly relevant, since $SO(16 \times SO(16)$ is not a subgroup of either of the heterotic gauge groups.  However, $w_2$ also governs the lifting of $[SO(16) \times SO(16)]/\Z_2$ bundles to $Spin(16) \bul Spin(16)$ bundles (see the diagram in Ref. [5]) and so, since $[SO(16) \times SO(16)]/\Z_2$ is contained in the projective special orthogonal  group $PSO(32)$, and the latter is the structural group of the Chan-Paton bundles of open string theory, $w_2$ is important in Type I theory (and possibly in establishing the $S$-duality of Type I with the heterotic $Semispin(32)$ theory [17].)
\bs
Notice that the vanishing of the obstructions $x_2$ and $y_2$ does not in itself suffice to establish T-duality.  If $Q_1$ and $Q_2$ are bundles with $x_2(Q_1) = y_2(Q_2) = 0$, then both $Q_1$ and $Q_2$ have non-trivial $(Spin(16) \ast Spin(16)) \triangleleft \Z_2$ double covers, but these covers {\it need not be the same}.  If they are different, then T-duality cannot be implemented.  This is why we insist that if $Q_1$ is a $(Spin(16) \bul Spin(16)) \triangleleft \Z_2$ bundle with \hfil\break $x_2(Q_2) = 0$, and $Q^\ast$ is the corresponding double cover, then $Q^\ast/\{ 1 \ast 1, {\hat K}_{16} \ast 1 \}$ is {\it the} T-dual $(Semispin(16) \times Semispin(16)) \triangleleft \Z_2$ bundle; for this bundle obviously does have $Q^\ast$ as
a double cover. (Obviously $y_2(Q^\ast/\{1 \ast 1, {\hat K}_{16} \ast 1\}) = 0$, so $x_2$ and $y_2$ are related in this limited sense.)
\bs
Finally, there can be additional complications if the {\it first} cohomology group $H^1(M,\Z_2)$ does not vanish.  It follows from  the second exact sequence above that if $Q_1$ is a $(Spin(16) \bul Spin(16)) \triangleleft \Z_2$ bundle with $x_2(Q_1) = 0$, then the double cover $Q^\ast$ need not be unique: the number of possibilities is counted by $|H^1(M,\Z_2)|$, the order of $H^1(M,\Z_2)$.  Strictly speaking, then, we must label the various $(Spin(16) \ast Spin(16)) \triangleleft \Z_2$ bundles $Q^\ast_i$, where $i = 1 \cdots |H^1(M,\Z_2)|$, corresponding to a {\it given} $Q_1$.  The winding modes might  then fall into topological sectors labelled by $i$.  (For toral compactifications, one might interpret these gauge spin structures in terms of boundary conditions.)  The problem now is that, in general, the $(Semispin(16) \times Semispin(16)) \triangleleft \Z_2$ bundles
$$Q^\ast_i/\{1 \ast 1, {\hat K}_{16} \ast 1 \}, i = 1 \cdots |H^1(M,\Z_2)|,$$
will not be mutually isomorphic.  This would mean that a certain given $Semispin(32)$ configuration might be ``T-dual" not to one  but rather to a whole {\it collection} of \hfil\break $(E_8 \times E_8) \triangleleft \Z_2$ configurations.  (Note that requiring $H^1(M,\Z_2) = 0$ would be very drastic, as it would rule out most manifolds with finite fundamental groups of even order.)  This situation can be avoided only by confining all winding modes to the same gauge spin bundle.
\bs
We may summarise as follows.  The fact that the ``$SO(16) \times SO(16)$" subgroups of the heterotic string gauge groups are {\it not} mutually isomorphic is not a problem in itself; indeed, it plays a key role in T-duality.  On the other hand, it also imposes topological conditions.  If we begin on the $Semispin(32)$ side, this condition is that the reduced gauge bundle $Q_1$ must admit an {\it exceptional structure}, which means that a certain cohomology class $x_2(Q_1)$, analogous to (but different from) the second Stiefel-Whitney class, must vanish.  If $x_2(Q_1) \not= 0$, then T-duality is topologically obstructed.
\bs
Let us consider some simple examples of these phenomena.
\bs
\nt
{\bf 6.  EXAMPLES}.
\bs
There are of course many ways of constructing principal fibre bundles exhibiting various kinds of topological non-triviality, but here we shall concentrate on three approaches which are particularly relevant to string theory.  All are straightforward and explicit.
\bs
\nt
\centerline{\bf A.  EMBEDDING THE SPIN CONNECTION IN THE GAUGE GROUP}
\bs
The ``traditional" method of constructing gauge vacua in string theory [18] is motivated by the need to cancel anomalies.  The curvature tensor of the compactification manifold is ``equated" to the gauge field strength, a procedure known as ``embedding the spin connection in the  gauge group".  (One should not be misled by the terminology : in the physics literature, the ``spin connection" usually means the components of the  Levi-Civit${\acute a}$ connection with respect to an orthonormal basis.  Globally, it could mean {\it either} the Levi-Civit${\acute a}$ connection form {\it or} its pull-back to a spin bundle over the bundle of orthonormal frames.)  Let us examine the use of this technique in constructing  various non-trivial $Semispin(32)$ and $E_8 \times E_8$ bundles.
\bs
An important string compactification manifold is the four-dimensional $K3$ space [19].  This manifold admits a Ricci-flat Riemannian metric.  Let $SO(K3)$ be a bundle of oriented orthonormal frames over $K3$ (which is orientable).  Then $SO(K3)$ is an $SO(4)$ bundle.  As $K3$ is a Ricci-flat K${\ddot a}$hler manifold, $SO(K3)$ can be reduced to an $SU(2)$ subbundle $SU(K3)$; however, $SU(K3)$ need not be stable under the action of the (finite, but possibly non-trivial) group of isometries of $K3$, so it is preferable to use $SO(K3)$.  This means that we should embed $SU(2)$ in the gauge group through some natural embedding of either  $SO(4)$ or (if indeed we wish to use  a {\it spin} connection in the true sense) of $Spin(4)$.
\bs
In fact, both $SO(4)$ and $Spin(4)$ have natural embeddings in $Spin(16) \bul Spin(16)$, as follows.  First, $Spin(4)$ is a subgroup of $Spin(16)$ in the obvious way, so the subgroup 
$$\{A \bul 1 \}, A \in Spin(4),$$
is isomorphic to $Spin(4)$.  On the other hand, since clearly $(-1) \bul (-1) = 1 \bul 1$, we see that
$$\{ A \bul A \}, A \in Spin(4),$$
is isomorphic to $SO(4)$.  Let us concentrate first on this latter case, and determine the structure of the covers of $SO(4)$ in $Spin(16) \bullet Spin(16)$ and $Spin(16) \ast Spin(16)$.  We have
$$(-1) \bullet (-1) = 1 \bullet 1,$$
$$(-1) \ast (-1) \not= 1 \ast 1.$$
Thus $\{A \bullet A \}$, the cover of $SO(4)$ in $Spin(16) \bullet Spin(16)$, is still (perhaps surprisingly) isomorphic to $SO(4)$, whereas $\{A \ast A \}$, the cover in $Spin(16) \ast Spin(16)$, is isomorphic to $Spin(4)$.  Now take $SO(K3)$, and extend it in the usual way to a $Spin(16) \bul Spin(16)$ bundle.  (We shall ignore the disconnected version, since $H^1(K3, \Z_2) = 0$ and so $x_1(Q) = 0$ for all $(Spin(16) \bul Spin(16)) \triangleleft \Z_2$ bundles over $K3$.)  That is, define
$$Q(SO(K3)) = [SO(K3) \times Spin(16) \bul Spin(16)]/SO(4),$$
with $SO(4)$ acting to the left on $Spin(16) \bul Spin(16)$ as usual.  We can define a $Semispin(32)$ bundle $P(SO(K3))$ in the same way.  The Levi-Civit$\acute a$ connection corresponding to a Ricci-flat metric on $K3$ can be regarded as a connection one-form on $SO(K3)$, and it pushes forward to a gauge field on $Q(SO(K3))$ and $P(SO(K3))$; the gauge field strength will coincide with the curvature form of $K3$; in short, we have ``embedded the spin (actually, the Levi-Civit$\acute a$) connection in the gauge group", $Semispin(32)$.
\bs
Now set
$${\tilde Q}(SO(K3)) = [SO(K3) \times Spin(16) \bullet Spin(16)]/SO(4).$$
Clearly ${\tilde Q}(SO(K3))$ is a $Spin(16) \bullet Spin(16)$ bundle over $K3$, and it can be extended to a $Spin(32)$ bundle.  Now $SO(4)$, as a subgroup of $Spin(32)$, does not contain ${\hat K}_{16} \bullet {\hat K}_{16} = {\hat K}_{32}$,  and so in fact
$${\tilde Q}(SO(K3))/\{1, {\hat K}_{32} \} = Q(SO(K3)).$$
That is, ${\tilde Q}(SO(K3))$ is a non-trivial double cover of $Q(SO(K3))$; it is a vector structure.  We have
$${\tilde w}_2(Q(SO(K3))) = 0.$$
Thus, ``embedding the Levi-Civit$\acute a$ connection is $Semispin(32)$" always yields a $Semispin(32)$ configuration with a vector structure.
\bs
The $K3$ manifold is a spin manifold, that is, $SO(K3)$ has a non-trivial double cover $Spin(K3)$ which is a $Spin(4)$ bundle over $K3$.  (This spin structure is unique).  As the $SO(4)$ subgroup of $Semispin(32)$ is covered by $Spin(4)$ in $Spin(16) \ast Spin(16)$, we can define
$$Q^\ast(SO(K3)) = [Spin(K3) \times Spin(16) \ast Spin(16)]/ Spin(4),$$
and this is a $Spin(16) \ast Spin(16)$ bundle over $K3$.  As $Spin(4)$ {\it does} contain $(-1) \ast (-1)$, we have
$$Q^\ast(SO(K3))/\{1 \ast 1, (-1) \ast (-1) \} = Q(SO(K3)),$$
so that $Q^\ast(SO(K3))$ is another non-trivial double cover of $Q(SO(K3))$.  Evidently
$$x_2(Q(SO(K3))) = 0,$$
that is, ``embedding the Levi-Civit$\acute a$ connection of $K3$ in $Semispin(32)$" yields a configuration with an exceptional structure.  Thus, this configuration is globally T-dual to an $E_8 \times E_8$ configuration
$$\eqalign{
Q^\ast(SO(K3))/\{1 \ast 1,  & {\hat K}_{16} \ast 1 \} =  \cr
&[Spin(K3) \times Semispin(16) \times Semispin(16)]/Spin(4).\cr}$$
\bs
An alternative procedure is to use, instead of the Levi-Civit$\acute a$ connection, the {\it spin connection} in the true sense $-$ that is, the pull-back of the Levi-Civit$\acute a$ connection to $Spin(K3)$.  Embedding $Spin(4)$ in $Spin(16) \bullet Spin(16)$ as $\{A \bul 1 \}$, we see that the covers of $Spin(4)$ in $Spin(16) \bullet Spin(16)$ and $Spin(16) \ast Spin(16)$ are both isomorphic to $Spin(4)$.  If therefore we define
$$Q(Spin(K3)) = [Spin(K3) \times Spin(16) \bul Spin(16)]/Spin(4),$$
then $Q(Spin(K3))$ is a principal $Spin(16) \bul Spin(16)$ bundle over $K3$ with non-trivial double covers
$$\eqalign{
{\tilde Q}(Spin(K3)) &= [Spin(K3) \times Spin(16) \bullet Spin(16)]/Spin(4), \cr
Q^\ast(Spin(K3)) &= [Spin(K3) \times Spin(16) \ast Spin(16)]/Spin(4), \cr}$$
and so $Q(Spin(K3))$ is another $Semispin(32)$ configuration with both vector and exceptional structures.  Notice that, although ``embedding the Levi-Civit$\acute a$ connection" and ``embedding the spin connection" both lead to $Semispin(32)$ configurations possessing both vector and exceptional structures, the two procedures are indeed quite different.  A useful way to show this is to note that the Levi-Civit$\acute a$ connection breaks $Spin(16) \bul Spin(16)$ to $[SU(2) \times Spin(12)] \bul [SU(2) \times Spin(12)]$ (since this is the centraliser of $SU(2)$ with this embedding), while the spin connection breaks it to \hfil\break $[SU(2) \times Spin(12)] \bul Spin(16)$.
\bs
Apart from $K3$, the only known examples of compact Ricci-flat Riemannian four-dimensional manifolds are the flat manifolds and the Enriques [20] and Hitchin [21] manifolds.  The Enriques manifold is a K$\ddot a$hler manifold of the form $K3/\Z_2$, while the Hitchin manifold is a non-K$\ddot a$hler manifold, $K3/[\Z_2 \times \Z_2]$.  Neither has attracted as much interest as $K3$, partly because neither is a spin manifold; see Ref. [22] for a discussion of this fact.  (As a general rule, it is difficult for compact, locally irreducible, Ricci-flat manifolds of dimension $n = 4r$ to be spin if they are not simply connected.  No example is known for $n = 4$ or $12$, and the only known examples for $n = 8$ have fundamental groups isomorphic to $\Z_2$.)  However, Pope et al [23] have argued that non-spin manifolds can be of interest in string theory when winding modes are taken into account; for example, the $AdS_5 \times S^1 \times \C P^2$ solution of IIA supergravity generates an acceptable BPS solution of the full string theory, despite the fact that $\C P^2$ is {\it not} a spin manifold.  Furthermore, the Enriques and Hitchin manifolds fail to be spin in a relatively innocuous way.  To explain this remark, we note first that, like all orientable four-dimensional manifolds [24], the Enriques manifold $K3/\Z_2$ and the Hitchin manifold $K3/(\Z_2 \times \Z_2)$ are both $Spin^c$ manifolds.  This essentially means that difficulties in constructing globally well-defined fermion fields can be overcome by coupling to a specific $U(1)$ ``gauge" field.  For the Enriques and Hitchin manifolds, this $U(1)$ field is particularly inconspicuous because its field strength {\it vanishes}; hence it does not affect the Lagrangian or the formula for the square of the Dirac operator.  In view of all this, we suggest that these manifolds may repay further investigation.
\bs
If we let $SO(K3/\Z_2)$ be a bundle of orthonormal frames over the Enriques manifold, then with the above embedding of $SO(4)$ in $Spin(16) \bul Spin(16)$ we can define
$$Q(SO(K3/\Z_2)) = [SO(K3/\Z_2) \times Spin(16) \bul Spin(16)]/SO(4).$$
One easily prove, in the same way as for $K3$, that 
$${\tilde w}_2 (Q(SO(K3/\Z_2))) = 0,$$
so we still have a vector structure in this case.  But if $Q(SO(K3/\Z_2))$ had a non-trivial $Spin(16) \ast Spin(16)$ cover, then $SO(K3/\Z_2)$ would have a non-trivial $Spin(4)$ double cover, and we know that this is not the case.  Hence
$$x_2(Q(SO(K3/\Z_2))) \not= 0,$$
and so $Q(SO(K3/\Z_2))$ is our first example of a $Semispin(32)$ configuration {\it with} a vector structure but  {\it without} an exceptional structure.  Precisely similar statements hold true of the Hitchin manifold:
$${\tilde w}_2(Q(SO(K3/(\Z_2 \times \Z_2)))) = 0,$$
$$x_2(Q(SO(K3/(\Z_2 \times \Z_2)))) \not= 0.$$
In short, ``embedding" the Levi-Civit$\acute a$ connections of these Ricci-flat manifolds in $Semispin(32)$ leads to configurations with no global T-dual partners.  This behaviour is best explained in terms of holonomy [25].  The linear holonomy group of the Hitchin manifold (endowed with a Yau metric descending [26] from the universal cover) is $(Q_8 \times SU(2))/\Z_2$, where $Q_8$ is the quaternion group of order $8$, and so this is the holonomy group of the gauge connection on $Q(SO(K3/\Z_2 \times \Z_2)))$.  If the latter had an exceptional structure, then the holonomy group of the pull-back connection would be $Q_8 \times SU(2)$.  However, it is impossible for a holonomy group to have eight connected components over a manifold with $\Z_2 \times \Z_2$ as fundamental group.  A similar argument explains the failure of T-duality for $Q(SO(K3/\Z_2))$.
\bs
T-duality can, however, be implemented on these manifolds if  we exploit their $Spin^c$ structures.  Recall [12] that oriented Riemannian manifold $M$ is said to have a $Spin^c$ structure if there exists a  $U(1)$ bundle $L(M)$ over $M$ such that $SO(M) + L(M)$ has a non-trivial double cover $Spin^c(M)$ which is a $Spin^c(n)$ bundle over  $M$.  Here $n = dim M$, $Spin^c(n) = (Spin(n) \times U(1))/\Z_2$, and  $SO(M) + L(M)$ is the submanifold of $SO(M) \times L(M)$ consisting of pairs $(s,t)$ such that $s$ and $t$ project to the same point in $M$.  (Thus $SO(M) + L(M)$ is an $SO(n) \times U(1)$ bundle over $M$; while $SO(M) \times L(M)$ is a bundle over $M \times M$.)  Clearly, $Spin^c$ structures are rarely unique when they exist, since there may be many choices for $L(M)$.  For K$\ddot a$hler manifolds, however, there is a natural choice : take a bundle of unitary frames $U(M)$, and set $L(M) = U(M)/SU(n/2)$, where $n$ is the real dimension of $M$.  For the Enriques manifold, which is K$\ddot a$hlerian, this $U(1)$ bundle inherits from $U(M)$ a connection with holonomy $\Z_2$.  The curvature is therefore zero, and so the $Spin^c$ structure can be employed without changing any {\it local} structures.  The Hitchin manifold is not K$\ddot a$hlerian, but for it, too, there is a canonical choice of $L(M)$, and again $L(M)$ has a flat connection, (Even with $L(M)$ fixed, however, there is still the usual source of ambiguity in defining $Spin^c(M)$, namely $H^1(M,\Z_2)$.)
\bs
Let us consider, then, the bundles $SO(K3/\Z_2) + L(K3/\Z_2)$ and \hfil\break $SO(K3/(\Z_2 \times \Z_2)) + L(K3/(\Z_2 \times \Z_2))$.  Both are $SO(4) \times SO(2)$ bundles over the respective base manifolds.  Now $SO(4) \times SO(2)$ is a subgroup of $SO(6)$, and the latter embeds naturally in
$Spin(16) \bul Spin(16)$, as $\{A \bul A, A \in Spin(6)\}$.  Hence we can define
$$\eqalign{
Q(SO(K3/\Z_2) & + L(K3/\Z_2)) \cr 
&= [(SO(K3/\Z_2) + L(K3/\Z_2)) \times Spin(16) \bul Spin(16)]/(SO(4) \times SO(2)),\cr}$$
and similarly for $Q(SO(K3/(\Z_2 \times  \Z_2))) + L(K3/(\Z_2 \times \Z_2)))$.  The Levi-Civit$\acute a$ connection form on $SO(K3/\Z_2)$, and the canonical flat connection  form on $L(K3/\Z_2)$, both pull back to $SO(K3/\Z_2) + L(K3/\Z_2)$, and their direct sum defines [1] a connection there.  This  connection pushes forward to a connection on the $Spin(16) \bul Spin(16)$ bundle $Q(SO(K3/\Z_2) + L(K3/\Z_2))$ and to the latter's $Semispin(32)$ extension.
\bs
It is easy to see that, since $SO(4) \times SO(2)$ is covered in $Spin(16) \bullet Spin(16)$ by a subgroup which is again isomorphic to $SO(4) \times  SO(2)$, $Q(SO(K3/\Z_2) + L(K3/\Z_2))$ and $Q(SO(K3/(\Z_2 \times \Z_2)) + L(K3/(\Z_2 \times \Z_2)))$ both have vector structures:
$${\tilde w}_2(Q(SO(K3/\Z_2) + L(K3/\Z_2))) = 0,$$
$${\tilde w}_2(Q(SO(K3/(\Z_2 \times \Z_2)) + L(K3/(\Z_2 \times \Z_2)))) = 0.$$
Now, however, let $Spin^c(K3/\Z_2)$ and $Spin^c(K3/(\Z_2 \times \Z_2))$ denote $Spin^c$ structures over the respective manifolds.  Then 
$$\eqalign{
Q^\ast(SO(K3/\Z_2)& + L(K3/\Z_2))  \cr
   &= [Spin^c(K3/\Z_2) \times Spin(16) \ast Spin(16)]/Spin(4) \bullet Spin(2)\cr}$$
is well-defined, $Spin(4) \bullet Spin(2)$ being the cover  of $SO(4) \times SO(2)$ in $Spin(16) \ast Spin(16)$, and similarly for $K3/(\Z_2 \times \Z_2)$.  (Through this discussion we have used the elementary isomorphisms [27]
$U(1) = SO(2) = Spin(2).)$.  That is,
$$\eqalign{
&x_2(Q(SO(K3/\Z_2) + L(K3/\Z_2))) = 0 \cr
&x_2(Q(SO(K3/(\Z_2 \times \Z_2)) + L(K3/(\Z_2 \times \Z_2)))) = 0; \cr}$$
both of these $Spin(16) \bul Spin(16)$ bundles have exceptional structures.  (The reader who wishes to investigate these configurations should note that while, for example, the holonomy groups of the connections on $SO(K3/\Z_2)$ and $L(K3/\Z_2)$ are isomorphic respectively to $[\Z_4 \times SU(2)]/\Z_2$ and $\Z_2$, the holonomy group of the direct sum connection on $SO(K3/\Z_2) + L(K3/\Z_2)$ is {\it not} $\Z_2 \times [\Z_4 \times SU(2)]/\Z_2$.  Rather, it is again isomorphic to $[\Z_4 \times SU(2)]/\Z_2$.  See Ref.[1], page 82.  The corresponding connection on $Spin^c(K3/\Z_2)$ {\it also} has holonomy $[\Z_4 \times SU(2)]/\Z_2$, {\it not} $\Z_4 \times SU(2)$.)  
\bs
We conclude, then, that if the Levi-Civit$\acute a$ connections of  the (Ricci-flat) Enriques and Hitchin manifolds are ``embedded" in $Semispin(32)$, we obtain configurations with no globally well-defined T-dual partners, even though both do have vector structures.  On the other hand, by exploiting the $Spin^c$ structures of these spaces, one can construct a different pair of  $Semispin(32)$ bundles which do admit T-dual partners.  It is interesting  to note here that the argument of Pope et al [23], to the effect that non-spin manifolds are acceptable in string compactifications, depends on the existence of $Spin^c$ structures and on the application of T-duality.
\bs
Six-dimensional Ricci-flat K$\ddot a$hler manifolds present fewer complications, because (in the compact case) these are  {\it always} spin whether they are simply connected or not.  Embedding $SO(6)$ in $Spin(16) \bul Spin(16)$ as $\{A \bul A, A \in Spin(6)\}$, we can extend $SO(CY)$, for a Calabi-Yau manifold $CY$, to a $Spin(16) \bul Spin(16)$ bundle $Q(SO(CY))$; such a bundle always has a vector structure, and, if $Spin(CY)$ is a spin structure over $CY$, we can use it to construct an exceptional structure.  Alternatively,  by embedding $Spin(6)$ in $Spin(16) \bul Spin(16)$ in the obvious way, we can use any spin structure over $CY$ to construct $Q(Spin(CY))$, which again has both vector and exceptional structures.  (As in the case of $K3$, $SO(CY)$ and $Spin(CY)$ are reducible bundles, but we avoid using the reduced bundles because they need not be mapped into themselves by isometries. Indeed, we should actually use a sub-bundle of the full bundle of orthonormal frames, $O(CY)$.  See Ref. [28] for a discussion of this point.)  We remind the reader that $Spin(CY)$ will not be unique if, as is frequently the case, $H^1(CY,\Z_2)$ does not vanish.  If more than one spin structure is  physically significant, then ``embedding" the Levi-Civit$\acute a$ or spin connections in $Semispin(32)$ will produce configurations which are ``T-dual" to a {\it family} of apparently distinct $(E_8 \times E_8) \triangleleft \Z_2$ configurations.  (Recall that we are using the term ``T-duality" in a very broad way in this work, to include {\it any} process of exchanging the gauge and matter fields of the two heterotic theories.  Such an exchange or comparison could be of interest even if T-duality in the more restricted sense (involving inversions of radii or other changes of moduli) cannot be implemented, which is apparently the case for Calabi-Yau compactifications [16]).
\bs
\centerline {\bf B. EXAMPLES FROM ABELIAN INSTANTONS.}
\bs
Gauge configurations with non-vanishing invariants such as ${\tilde w}_2$ or $x_2$ are of course topologically non-trivial.  One of the simplest but physically most relevant ways to construct such fields is to use the non-trivial $U(1)$ bundles over the  two-sphere (``Abelian instantons").  These arise naturally when the singularities of orbifolds are blown up [6].  Throughout this section, the base manifold is either a two-cycle in some manifold, or  a two-sphere around an orbifold singularity.  The construction is guided by the discussions in sections $4$ and $5$ of Ref. [6].
\bs
Recall that $SO(16)$ contains the unitary group $U(8)$, which is globally isomorphic to 
$[U(1) \times SU(8)]/ {\Z}_8$.  The corresponding subgroup of $Spin(16)$ is isomorphic [5] to  $[U(1) \times SU(8)]/ {\Z}_4$.  We denote the elements of the latter by $[u,s]_4$, so that $[iu,s]_4 = [u,is]_4$ and so on.  Now let ${\hat J}_8$ be the element of $Spin(16)$ defined by
$${\hat J}_8 = [i, I_8]_4;$$
\ss
\nt it projects to the $SO(16)$ matrix $\left[ \matrix{0& -I_8 \cr
                                                    &   \cr
                                                   I_8& 0 \cr} \right]$   
where $I_8$ is the $8 \times 8$ identity matrix.  One can show [5] that $({\hat J}_8)^2 = {\hat K}_{16}$.  Notice that ${\hat J}_8$ is contained in the $U(1)$ subgroup
$$U(1)_J = \{[u,I_8]_4, u \in U(1)\}.$$
Next, let $z$ be a primitive sixteenth root of unity, so that the matrix $diag (-z, z, z, z, z, z, z, z)$ is an element of $SU(8)$, and so 
$${\hat L}_8 = [z^{-1}, diag (-z, z, z, z, z, z, z, z)]_4$$
is an element of $[U(1) \times SU(8)]/ \Z_4$.  Notice that $({\hat L}_8)^2$ is $[z^{-2}, z^2I_8]_4$, which projects to $I_8$ in $U(8)$, and so $({\hat L}_8)^2 = -1 \in Spin(16)$.  Clearly ${\hat L}_8$ is contained in
$$U(1)_L = \{[u^{-1}, diag(u^{-7}, u, u, u, u, u, u, u)]_4, u \in U(1)\}.$$
Thus, $U(1)_J$ and $U(1)_L$ are {\it commuting} $U(1)$ subgroups of $[U(1) \times SU(8)]/ {\Z}_4$ and hence of $Spin(16)$.  If we define
$$\Delta(U(1)_J {\bul} U(1)_J) = \{u {\bul} u \in Spin(16) {\bul} Spin(16), u \in U(1)_J \},$$
$$\Delta (U(1)_L {\bul} U(1)_L) = \{u {\bul} u \in Spin(16) {\bul} Spin(16), u \in U(1)_L \},$$
then we obtain a pair of commuting $U(1)$ subgroups of $Semispin(32)$.  Let $H_J$, $H_L$ be bundles over the (same) two-sphere, both  isomorphic to the Hopf bundle [29], and with $\Delta(U(1)_J {\bul} U(1)_J)$ and $\Delta(U(1)_L {\bul} U(1)_L)$, respectively, as structural groups.  Note that, because $\Delta (U(1)_J {\bul} U(1)_J)$ and $\Delta (U(1)_L {\bul} U(1)_L)$ commute, $H_J + H_L$ is a well-defined $U(1) \times U(1)$ bundle.  Let $Q(H_J)$, $Q(H_L)$, and $Q(H_J + H_L)$ be the $Spin(16) {\bul} Spin(16)$ extensions of these bundles.  Then $Q(H_J)$ is essentially the bundle {\it without} a vector structure discussed in section 4 of Ref.[6].  This bundle does have an exceptional structure, however, as we shall soon prove; so T-duality is valid for $Q(H_J)$.
\bs
Let us consider $Q(H_L)$.  If $G$ is any Lie group with a non-trivial double cover then the pre-image of any $U(1)$ in $G$ can be isomorphic either to $U(1)$ or to $\Z_2 \times U(1)$.  Since the non-trivial element of the kernel of the projection $Spin(16) \ast Spin(16) \to Spin(16) \bul Spin(16)$ is $(-1) \ast (-1)$, we see that this projection induces a non-trivial covering of $\Delta(U(1)_L {\bul} U(1)_L)$ if and only if $-1 \in U(1)_L$, which is indeed the case since $U(1)_L$ contains ${\hat L}_8$ and
$({\hat L}_8)^2 = -1$.  Now it follows that if $x_2(Q(H_L)) = 0$, so that $Q(H_L)$ has a non-trivial (double) $Spin(16) \ast Spin(16)$ cover, then $H_L$ has a non-trivial double cover.  But this is not the case; the Hopf bundle has no square root [29].  Thus, in fact, $Q(H_L)$ has no exceptional structure : T-duality is topologically obstructed for this $Semispin(32)$ gauge configuration.  On the other hand, since $U(1)$ has a {\it unique} element of order 2, we see that $U(1)_L$ does not contain ${\hat K}_{16}$, and so, since ${\hat K}_{16} \bullet {\hat K}_{16}$ is the non-trivial element of the kernel of the projection $Spin(16) \bullet Spin(16) \to Spin(16) \bul Spin(16)$, the pre-image of $\Delta(U(1)_L {\bul} U(1)_L)$ in $Spin(16) \bullet Spin(16)$ is isomorphic to $\Z_2 \times U(1)$.  We can therefore construct a $Spin(16) \bullet Spin(16)$ non-trivial double cover of $Q(H_L)$, despite the fact that  $H_L$ has no square root.  Thus $Q(H_L)$ has a vector structure.  Arguing in the same way, one shows that the opposite statements are true of $Q(H_J)$; for $U(1)_J$ contains ${\hat J}_8$ and therefore ${\hat K}_{16} \left( = \left({\hat J}_8\right)^2\right)$, while it does not contain $-1$.
\bs
Finally, notice that if $\left(h_J, h_L\right)$ is any element of $H_J + H_L$, then the maps $(h_J, h_L) \to h_J$, $(h_J, h_L) \to h_L$ are bundle homomorphisms onto $H_J$ and $H_L$ respectively, and so coverings of $Q(H_J + H_L)$ yield coverings of $Q(H_J)$ and $Q(H_L)$.  Thus if $Q(H_J + H_L)$ had a vector structure, so would $Q(H_J)$, and if  $Q(H_J + H_L)$ had an exceptional structure, so would  $Q(H_L)$.  Hence in fact $Q(H_J + H_L)$ yields an example of a $Semispin(32)$ gauge configuration with {\it neither} a vector structure {\it nor} an exceptional structure.  In summary, we have 
$${\tilde w}_2(Q(H_J)) \not= 0; \;x_2(Q(H_J)) = 0;$$
$${\tilde w}_2(Q(H_L)) = 0;\; x_2(Q(H_L)) \not= 0;$$
$${\tilde w}_2(Q(H_J + H_L)) \not= 0; \;x_2(Q(H_J + H_L)) \not= 0.$$
In a similar way, let $U(1)_E$ be a $U(1)$ subgroup of $Semispin(16)$ containing $J^\ast_8$, the projection of ${\hat J}_8$, and let  $H_E$ be a Hopf bundle with $U(1)_E$ as structural group.  Extend $H_E$ to a $Semispin(16) \times Semispin(16)$ bundle $Q(H_E)$.  Then the pre-image of $U(1)_E$ in $Spin(16) \ast Spin(16)$ must be isomorphic to $U(1)$, since $\left(J_8^\ast, 1\right)$ is covered by ${\hat J}_8 \ast 1$, which squares to ${\hat K}_{16} \ast 1$, the non-trivial element of the kernel of $Spin(16) \ast Spin(16) \to Semispin(16) \times Semispin(16)$.  We conclude that $Q(H_E)$ is an $E_8 \times E_8$ configuration with  no T-dual partner; that is, $y_2(Q(H_E)) \not= 0$.
\bs
\centerline{\bf C. EXAMPLES INVOLVING ORBIFOLDS.}
\bs
T-duality was, of course, originally defined with respect to tori; however, one can attempt to extend it to other manifolds (such as K3) by regarding them as desingularisations of orbifolds.  It is shown in Ref.[6] that Abelian instantons can ``hide" in the singularities of $T^4/\Z_2$, where $T^4$ is the four-torus.  By blowing up the singularities, one can if necessary bring the hidden instantons into the open, and the techniques of the preceding section can be applied.  The most striking examples of this kind have a hidden instanton which is a $Semispin(32)$ configuration similar to $Q(H_J)$ above : it has no vector structure, but it does have an exceptional structure, and it is shown in Ref. [6] that it is T-dual to the $E_8 \times E_8$ DMW vacuum [15].  (In fact, the DMW vacuum is completely symmetric with respect to the two $E_8$ factors, and the gauge group is actually $(E_8 \times E_8) \triangleleft \Z_2$, so this is a case where (see Section 5 above) $y_1 \not= 0$.)
\bs
Other vacua considered in Ref.[6] do have vector structures, but some of these do {\it not} have exceptional structures.  Before discussing these, however, we must  clarify the following point, first mentioned in Ref. [6].  When discussing T-duality on $\R^9 \times S^1$, one often begins with unbroken $Semispin(32)$, and ``continuously turns on" a Wilson line that breaks $Semispin(32)$ to $(Spin(16) \bul Spin(16)) \triangleleft \Z_2$.  Then T-duality converts this to a $(Semispin(16) \times Semispin(16)) \triangleleft \Z_2$ theory, which becomes an $(E_8 \times E_8) \triangleleft \Z_2$ theory when the Wilson line is ``continuously turned off".  However, it is intuitively clear that if  we insist that it should  always be possible to turn Wilson lines on and off ``continuously", then we shall exclude topologically non-trivial configurations such as $Semispin(32)$ bundles lacking vector structures.  To see this, let $P$ be a $Semispin(32)$ bundle with a family of gauge connections, $\{\Gamma_t, t \in [0,1]\}$.  Suppose that, for each $t$, the holonomy group of $\Gamma_t$ is a cyclic group (typically of infinite order) generated by an element $W^\ast_t \in Semispin(32)$, such that 
$$\eqalign{
W^\ast_0 &= 1, \cr
W^\ast_1 &= {\hat K}_{16} {\bul} 1.\cr}$$
Then, since $(Spin(16) \bul Spin(16)) \triangleleft \Z_2$ is the centraliser of ${\hat K}_{16} {\bul} 1$, we have here an explicit realisation of ``turning the Wilson line on continuously".  But the ``reduction theorem" [1] states that $P$ is reducible to a holonomy bundle of {\it any} connection on $P$.  For $\Gamma_0$, such a holonomy bundle is a cross-section, and so $P$ {\it is trivial}.  As a trivial $Semispin(32)$ bundle always has vector and exceptional structures, and as the DMW vacua are undoubtedly dual to $Semispin(32)$ configurations without vector structures, we are forced to abandon the notion that  duality necessarily involves ``continuous turning on" of Wilson lines.  This is a natural consequence of permitting compactifications more complicated than $S^1 \times \R^9$.
\bs
In fact, a more physical argument to this effect is advanced in Ref.[6].  Henceforth, therefore, we shall {\it not} require $Semispin(32)$ to be broken to $(Spin(16) \bul Spin(16)) \triangleleft \Z_2$ by a Wilson line that can be continuously turned on; in fact, we should allow {\it any} $Semispin(32)$ bundle admitting a $(Spin(16) \bul Spin(16)) \triangleleft \Z_2$ subbundle.  This allows us to accept vacua without vector structures, but, at the same time, it forces us to deal with vacua lacking {\it exceptional} structures.  In the orbifold context, these arise, in a very remarkable way, from the detailed structure of the ``twisting matrix".
\bs
If a given $Semispin(32)$ configuration {\it does} have a vector structure, then we can use
the {\bf 32} representation of $SO(32)$, and the  theory may be constructed using $32$ left-moving fermions in such a multiplet.  If we wish  to study T-duality on the orbifold $T^4/{\Z}_2$, then, in projecting from $T^4$, we must  effect a ``twist" in the gauge group in order to preserve level matching [30].  The twist acts  on {\bf 32} by the $SO(32)$ matrix $X(m,n)$, a diagonal matrix with non-zero entries equal to $\pm 1$, with $m$ negative entries in the first $16$ places and $n$ negative entries in the last $16$ places, $m + n$ being even.  Level matching requires $m + n$ to be $4$ modulo $8$.  We shall consider the case $m + n = 12$.  Here  $Semispin (32)$ is broken to $Spin(20) \bul Spin(12)$, the double dot meaning as usual that the centres (both isomorphic to $\Z_2 \times \Z_2)$ are completely identified.
\bs
Let us consider first the simplest case, the case in which the Wilson line can be \hfil\break `` turned on continuously".  Such a configuration is topologically trivial, and should have both vector and exceptional structures : let us confirm this.  Let $W^\ast _t$ be the holonomy element defined  earlier, and let $W_t$ be the corresponding $SO(32)$ matrix.  Now $\Z_2$ acts on $T^4$ by reversing the signs of all four ``Cartesian" coordinates.  It therefore reverses the sense in which a non-contractible loop is traversed.  If $W_t$ is to descend to $T^4/{\Z}_2$,  therefore, we must have
$$X(m,n)W_tX(m,n)^{-1} = W_t^{-1}, \hbox{for all}\; t.$$
Since $W_1 = {\rm diag}(-I_{16}, I_{16})$, we can set $W_t = {\rm exp}(\pi t b)$, where, following Ref.[6], we put
$$b = \left[ \matrix{ 0&  -I_8& \cr
                      I_8&  0&  \cr
                       &  &  0_{16}\cr}\right].$$
Thus we need $X(m,n) b X(m,n)^{-1} = -b$, and so in this case $m = 8$, and so $n = 4$.  Thus $X(8,4)$ is the twisting matrix {\it when the Wilson line is ``continuously turned on"}.
\bs
Now the elements of $Spin(32)$ and $Semispin(32)$ corresponding to  $X(8,4)$ are, respectively, ${\hat K}_8 \bullet {\hat K}_4$ and ${\hat K}_8 {\bul} {\hat K}_4$, and the relevant element of $Spin(16) \ast Spin(16)$ is  ${\hat K}_8 \ast {\hat K}_4$.  As $({\hat K}_8)^2 = ({\hat K}_4)^2 = 1$, we see that ${\hat K}_8 \bullet {\hat K}_4$, ${\hat K}_8 \ast {\hat K}_4$, and ${\hat K}_8 {\bul} {\hat K}_4$ are all of order two.  Any $U(1)$ subgroup of $Spin(16) {\bul} Spin(16)$ which contains ${\hat K}_8 {\bul} {\hat K}_4$ is therefore  covered in $Spin(16) \bullet Spin(16)$ and $Spin(16) \ast Spin(16)$ by groups isomorphic to $\Z_2 \times U(1)$.  It follows that if such a $U(1)$ is the structural group of a $U(1)$ bundle which extends to an $Spin(16) \bul Spin(16)$ bundle, then the latter has non-trivial $Spin(16) \bullet Spin(16)$ and $Spin(16) \ast Spin(16)$ double covers.  As expected, this $Semispin(32)$ configuration has both vector and exceptional structures, and T-duality can be implemented in this case.
\bs
If, however, we drop the requirement that Wilson lines be ``continuously turned on", then we can allow other twisting matrices $X(m,n)$, with $m + n = 12$.  For example, we might take $X(10,2)$.  Now $({\hat K}_{10})^2 = ({\hat K}_{2})^2 = -1$, but this is not a problem in $Semispin(32)$, since we have
$$(({\hat K}_{10}) {\bul} ({\hat K}_{2}))^2 = (-1) {\bul} (-1) = 1,$$
so the twisting element is still of order two, as it must be  for a $\Z_2$ orbifold.  Similarly, $(-1) \bullet (-1) = 1$ in  $Spin(16) \bullet Spin(16)$ $-$ that is, in $Spin(32)$ $-$ so $({\hat K}_{10}) \bullet ({\hat K}_{2})$ is likewise of order two.  From our earlier discussions, we conclude that the model with the $X(10,2)$ twisting matrix still has a vector structure.  But we have
$$(({\hat K}_{10}) \ast ({\hat K}_{2}))^2 = (-1) \ast (-1) \not= 1$$
in $Spin(16) \ast Spin(16)$.  If we try to lift to $Spin(16) \ast Spin(16)$, we will have a twisting element of order four, which does not make sense on a $\Z_2$ orbifold.  This is a signal that {\it there is no exceptional structure in this case}.  Similarly, the matrix $X(6,6)$ corresponds to a model {\it with} a vector structure but {\it without} an exceptional structure.
\bs
It may perhaps seem odd that $X(8,4)$ and $X(10,2)$  can lead to distinct models, since they appear to represent different ways of distributing 12 minus signs in a diagonal $SO(32)$ matrix.  However, ${\hat K}_{8} \ast {\hat K}_{4}$ and ${\hat K}_{10} \ast {\hat K}_{2}$ are certainly very different : they are not mutually conjugate, for example.  The point to bear in mind is that, unlike $Spin(16) \bullet Spin(16)$, 
$Spin(16) \bul Spin(16)$, and $SO(16) \times SO(16)$, the group $Spin(16) \ast Spin(16)$ is {\it not} contained in any group locally isomorphic to  $SO(32)$.  Consequently, two apparently very similar $SO(32)$ or $Semispin(32)$ configurations can behave very differently with respect to T-duality.  In this instance, the precise  form of the twisting matrix determines whether a global T-dual partner exists at all.
\bs
\nt {\bf 7.  CONCLUSION}
\bs
The principal  findings of this work can be stated very simply.  On $\R^9 \times  S^1$, the two heterotic string theories are related by T-duality : one theory compactified on a circle of radius $R$ is equivalent to the other compactified on a circle of radius proportional to $1/R$.  As soon, however, as one goes to manifolds of greater complexity, the analogous statement is questionable : T-duality can be obstructed topologically.  This is true even on so simple a manifold as $T^2 \times \R^8$, even though one may wish to invert the radius of only one circle.  That is, whether T-duality works for a given circle depends on the context of that circle.  (See Ref. [16] for a much more subtle instance of this.)  More mundanely, the essential point here is that the two gauge groups, $E_8 \times E_8$  and $Semispin(32)$, are not quite as similar as their Lie algebras might lead one to expect.
\bs
From a practical point of view, our results mean that any discussion of the relationship between the two heterotic theories must involve a computation of the ``exceptional Stiefel-Whitney class"  $x_2$ (or of $y_2$ if one begins on the $E_8 \times E_8$ side).  If $x_2$ fails to vanish, then a comparison is not meaningful, whatever the local situation may suggest.
\vfil\eject
\i {\bf REFERENCES}/

\bs

\i [1]  /S. Kobayashi and K. Nomizu, Foundations of Differential Geometry Vol. I, Interscience, New York, 1963.

\i [2]  /B. McInnes, J. Math. Phys. {\bf 38} (1997) 4354.

\i [3]  /J. Polchinski, String Theory, Cambridge University Press, New York, 1998.

\i [4]  /W. Lerche, C. Schweigert, R. Minasian, and S. Theisen, Phys. Lett. {\bf B424} (1998) 53.

\i [5]  /B. McInnes, The Semispin Groups in String Theory, to appear in J. Math. Phys., hep-th/9906059.

\i [6]  /M. Berkooz, R. G. Leigh, J. Polchinski, J. H. Schwarz, N. Seiberg, and E. Witten, Nucl. Phys. {\bf B475} (1996) 115.

\i [7]  /J. H. Schwarz, Nucl. Phys. Proc. Suppl. {\bf 55B} (1997) 1.

\i [8]  /D. Husemoller, Fibre Bundles, Springer-Verlag, Berlin, 1975.

\i [9] /T. Br$\ddot o$cker and T  tomDieck, Representations of Compact Lie Groups, Springer-Verlag, Berlin, 1985.

\i [10] /G. G. Ross, Grand Unified Theories, Addison-Wesley, Reading, 1984.

\i [11]  /L. O'Raifeartaigh, Group Structure of Gauge Theories, Cambridge University Press, Cambridge, 1987.

\i [12]  /H. B. Lawson and M. L. Michelsohn, Spin Geometry, Princeton University Press, Princeton, 1989.

\i [13]  /A. Borel, Tohoku Math. Jour. {\bf 13}(1962) 216.

\i [14]  /S. Chaudhuri, G. Hockney, and J. Lykken, Phys. Rev. Lett, {\bf 75}(1995)2264.

\i [15]  /M. J. Duff, R. Minasian, and E. Witten, Nucl. Phys. {\bf B465} (1996) 413.

\i [16]  /P. S. Aspinwall and M. R. Plesser, T-Duality Can Fail, hep-th/9905036.

\i [17]  /E. Witten, J.H.E.P. {\bf 9812} (1998) 019.

\i [18]  /M. B. Green, J. H. Schwarz, and E. Witten, Superstring Theory, Cambridge University Press, Cambridge, 1987.

\i [19]  /P. S. Aspinwall, K3 Surfaces and String Duality, hep-th/9611137.

\i [20]  /W. Barth, C. Peters and A. van de Ven, Compact Complex Surfaces, Springer-Verlag, Berlin, 1984.

\i [21]  /N. Hitchin, J. Diff. Geom. {\bf 9}(1974) 435.

\i [22]  /B. McInnes, Commun. Math. Phys. {\bf 203} (1999)  349.

\i [23]  /C. N. Pope, A. Sadrzadeh, and S. R. Scuro, Timelike Hopf Duality and Type IIA$^\ast$ String Solutions, hep-th/9905161.

\i [24]  /J. W. Morgan, The Seiberg-Witten Equations and Applications to the Topology of Smooth Four-Manifolds, Princeton University Press, Princeton, 1996.

\i [25]  /A. L. Besse, Einstein Manifolds, Springer-Verlag, Berlin, 1987.

\i [26]  /B. McInnes, J. Math. Phys. {\bf 34} (1993) 4287.

\i [27]  /M. L. Curtis, Matrix Groups, Springer-Verlag, Berlin, 1984.

\i [28]  /B. McInnes, J. Math. Phys. {\bf 40} (1999) 1255.

\i [29]  /R. Bott and L. W. Tu, Differential Forms in Algebraic Topology, Springer-Verlag, Berlin, 1982.

\i [30]  /L. Dixon, J. Harvey, C. Vafa, and E. Witten, Nucl. Phys. {\bf B274} (1986)285.
\bs

\bs
\bye

hys. B274(1986)285.
\bye